\documentclass[12pt]{article}
\usepackage{amsfonts,latexsym,amssymb}
\usepackage[english]{babel}
\usepackage[dvips]{graphicx,color}
\usepackage{amsmath}
\usepackage[mathscr]{eucal}
\usepackage{rotating}
\newcommand\Aamp[2]{A_{#1}^{\;\;\;#2}}
\newcommand\Aampk[3]{A_{#3#1}^{\;\;\;\;#2}}

\def\be{\begin{equation}}

\def\blue          {\color{mydarkblue}}

\def\CA{\mathcal{A}}

\newcommand\CMt[1]  {\M_{\chi;#1}}
\newcommand\CMtd[1]  {\M^*_{\chi;#1}}
\def\C{\ensuremath{\mathcal C}}
\newcommand\coronebd[2]{c(#1;#2)}
\def\Corr{\text{Corr}}

\def\dim{\text{dim}}

\def\ee{\end{equation}}
\def\End{\text{End}}
\def\eps{\varepsilon}
\newcommand\eqpic[4]{\begin{eqnarray}
                   \begin{picture}(#2,#3){}\end{picture}\nonumber\\
                   \raisebox{-#3pt}{ \begin{picture}(#2,#3) #4 \end{picture} }
                   \label{#1} \\~\nonumber \end{eqnarray} }

\def\gree          {\color{ForestGreen}}

\def\Hom{\ensuremath{\text{Hom}}}
\newcommand\Homa[2]{\ensuremath{\text{Hom}_{A}(#1,#2)}}
\newcommand\Homaa[2]{\ensuremath{\text{Hom}_{A|A}(#1,#2)}}
\newcommand\Homab[2]{\ensuremath{\text{Hom}_{A|B}(#1,#2)}}

\def\I{\ensuremath{\mathcal I}}

\newcommand\Includepic[1]   {{\begin{picture}(0,0)(0,0)
                            \scalebox{.38}{\includegraphics{imgs/#1.eps}}\end{picture}}}
\def\id		                {\text{id}}
\def\inv                    {^{-1}}

\def\kb                     {\ensuremath {{\bar k}}}

\def\longrsqarrow  {\scalebox{.38}{\includegraphics{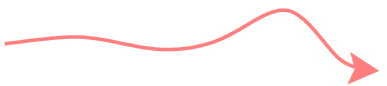}}}

\def\lsqarrow      {\scalebox{.38}{\includegraphics{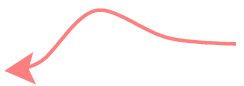}}}

\def\M                     {\ensuremath{{\mathscr M}}}
\newcommand\MAnnk[6]       { (\mathscr{A}_{#3#1}^{\;\;\;#2})_{#4#5#6}}
\def\Mfg                  {\ensuremath{{\mathscr {\tilde M}}_{\rm {X'}}}}

\def\MX                     {\ensuremath{{\mathscr M}_{\rm X}}}
\def\MXcut                     {\ensuremath{{\mathscr M}^\circ_{\rm X}}}
\def\MXf                     {\ensuremath{{\mathscr { M}}_{\rm {X'}}}}

\def\millstone                    {\ensuremath{{\mathscr N}}}

\newcommand\nbf[4]{\ensuremath{\nbfm_{#1#2,#3#4}}}
\newcommand\nbfinv[4]{\ensuremath{(\nbfm\inv)_{#1#2,#3#4}}}
\def\nbfm{\omega}

\newcommand\N[3]            {\mathcal{N}_{#1#2}^{\;\;\;#3}}

\def\one           {{\bf1}}
\def\oti{{\otimes}}

\newcommand\pb[1]   {\sse$\blue #1$}
\newcommand\pg[1]   {\sse$\gree #1$}
\newcommand\pl[1]   {\sse$#1$}

\def\qb{\bar q}

\def\Rep{\mathcal{R}ep}
\newcommand\refc[3]{b^{#1,#2}_{#3}}

\def\sse           {\scriptsize}

\def\T             {{\mathscr T}}
\def\threecob{\ensuremath{3{\text{-}}\mathcal{C}ob}}
\def\tfm{Z}
\def\tfs{\mathcal{H}}
\def\tft{\ensuremath{\text{\ttfamily{tft}}_{\C}}}
\newcommand\torchar[1]{|\chi_{#1};T\rangle}
\newcommand\torchard[1]{\langle\chi_{#1};T|}

\def\v{^\vee}
\def\V{\ensuremath{\mathfrak{V}}}
\def\Vect{\mathcal Vect_{\mathbb C}}
\def\vio           {\color{DarkViolet}}

\def\WS{X}
\def\wsemb{\iota}
\def\WSf{X'}
\def\WSg{X^c}

\definecolor{DarkViolet} {rgb}  {0.580392,0.000000,0.827450}
\definecolor{ForestGreen}{rgb}  {0.100000,0.408823,0.100000}
\definecolor{mydarkblue} {rgb}  {0.282352,0.239215,0.803921}

\begin{document}


\begin{center}\Large
Factorization constraints  and boundary conditions in rational CFT
\end{center}\vskip 2.1em

\begin{center}
  ~Carl Stigner\,
\end{center}
\begin{center}\it
  Teoretisk fysik, \ Karlstads Universitet\\
  Universitetsgatan 21, \,651\,88\, Karlstad
\end{center}

\begin{center}
  ~June 30 2010
\end{center}

\begin{abstract}
Among (conformal) quantum field theories, the rational conformal field theories are singled out by the fact that their correlators can be constructed from a modular tensor category $\C$ with a distinguished object, a symmetric special Frobenius algebra $A$ in $\C$, via the so-called TFT-construction. These correlators satisfy in particular all factorization constraints, which involve gluing homomorphisms relating correlators of world sheets of different topology.

We review the action of the gluing homomorphisms and discuss the implications of the factorization constraints for boundary conditions. The so-called classifying algebra $\CA$ for a RCFT is a semisimple commutative associative complex algebra, which classifies the boundary conditions of the theory. We show that the annulus partition functions can be obtained from the representation theory of $\CA$.
\end{abstract}

\section{Introduction}
There are various physical motivations to study quantum field theories on two-dimensional compact manifolds with a complex structure, possibly with non-empty boundary. Applications appear e.g.\! in condensed matter physics and in string theory. Such a surface is, by terminology inherited from string theory, called a \emph{world sheet}.
The situation becomes particularly interesting for a (full, local) conformal field theory (CFT), i.e. a two-dimensional QFT with conformal symmetry defined on world sheets. In two dimensions, there are, apart from the global conformal transformations, an infinite number of local conformal transformations giving rise to an infinite dimensional symmetry algebra.
In fact, as a consequence of the huge amount of symmetry, conformal field theories can be studied in a fully non-perturbative manner. This is another reason to study 2d CFT.

We denote by $\WSg$ a world sheet, possibly with boundary, and a number of field insertions in the bulk or on the boundary.
The correlation function $\Corr(\WSg)$ for the world sheet $\WSg$ associates to $\WSg$ a map from the relevant space of fields to the complex numbers. Correlation functions are linear in the fields and satisfy a number of consistency conditions. Among them are the factorization constraints, which can be thought of as a concrete realization of the notion of inserting a complete set of states. Solving a CFT amounts to giving the correlation function for any world sheet $\WSg$. This paper concerns a special class of CFT's, the so-called \emph{rational} CFT's (RCFT), for which there is a nice description of the construction in terms of modular tensor categories.

An important issue in CFT is the classification of conformal boundary conditions. A priori this is a difficult problem, except for some simple models. In e.g.\! the Ising model, a simple spin model, all boundary conditions  can be described in terms of a fixed external magnetic field applied to the spin variables at the boundary. This gives rise to a one parameter family of boundary conditions, which renormalize to three different boundary conditions in the continuum model. Two of them, spin up and spin down, correspond to a non-zero external magnetic field, whereas the third one, the free boundary condition, corresponds to taking the external magnetic field to be zero. However, it is far from obvious that boundary conditions of this form exhaust the conformal boundary conditions. E.g.\! in the three-states Potts model, there is one conformal boundary condition which can not be related in a simple way to the external magnetic field \cite{AOS}.

In \cite{FuS1} it was conjectured that the conformal boundary conditions for a specific class of theories are classified by a semi-simple commutative associative complex algebra, the so-called classifying algebra $\CA$.
In \cite{FSS} we establish the existence of the classifying algebra for any RCFT. The structure constants of $\CA$ are obtained by comparing bulk and boundary factorization of a disc with two bulk field insertions. The irreducible representations of $\CA$ are the so-called reflection coefficients. The reflection coefficients, which appear e.g.\! in \cite{CaLe,FuS1}, are collected in so-called boundary states. The boundary states contain essential physical information regarding boundary conditions, such as ground state degeneracies \cite{AL} and Ramond-Ramond charges of string compactifications \cite{BDLR}.
Moreover it has been shown, for some special classes of models, see e.g.\! \cite{BPPZ,Ca,PSS}, that the reflection coefficients appears naturally in the annulus partition functions. In this paper we show, by applying bulk factorization, that essential information concerning the annulus partition functions for any RCFT is contained in  $\CA$ and its representation theory. Thus the appearance of the reflection coefficients in the annulus coefficients of a RCFT is a generic phenomenon.

The symmetries of a CFT can be encoded in the mathematical structure of a conformal vertex algebra $\V$, by physicists often referred to as the chiral algebra.
A rational CFT is distinguished by the property that the strictification of the category $\Rep(\V)$ of representations of $\V$ is a modular tensor category $\C$. The correlation functions of a rational CFT satisfy holomorphic factorization, e.g.\! the correlation function $\Corr(\WSg)$ is a vector in the space of conformal blocks on the double $\widehat\WSg$. The double is  obtained from $\WSg$ by taking the orientation bundle over $\WSg$ and pairwise identify points over the boundary $\partial\WSg$:
\be\label{double}
    \widehat \WSg:=\text{or}(\WSg)\,\big/\sim\;,\quad(x,\text{or})\sim(x,-\text{or})\;\forall x\in\partial \WSg.
\ee
 The double is in particular a complex curve, thus we can study the space of conformal blocks on $\widehat\WSg$.

The solution of a rational conformal field theory, with given chiral algebra $\V$, can be split off into two separate parts, a complex-analytic and a purely algebraic part. The first problem amounts to solving the chiral theory on $\widehat\WSg$, i.e. to obtain the space of conformal blocks of $\V$ on $\widehat\WSg$. The second problem amounts to selecting, from the space of conformal blocks, the particular vector $\Corr(\WSg)$. This paper is concerned with the second problem. As a consequence, we will be able to restrict to topological world sheets. A topological world sheet $\WS$ is obtained from $\WSg$ by suppressing the conformal structure.

This paper is formulated in the framework of the TFT-construction. The TFT-construction provides all solutions to a rational CFT with given chiral algebra $\V$. A rational CFT, with chiral algebra $\V$, is constructed from the modular tensor category $\C$, which is the strictification of $\Rep(\V)$, and a distinguished object $A$ in $\C$, with the structure of a symmetric special Frobenius algebra. In fact, the rational CFT's with chiral algebra $\V$ are classified by Morita classes of simple symmetric special Frobenius algebras in $\C$.
We will not discuss vertex algebras explicitly, we will rather work in the framework of an abstract modular tensor category. Thereby we cover all rational CFT's simultaneously.

A crucial tool in the TFT-construction is a topological field theory. A topological field theory is a tensor functor $\tft$ from the category $\threecob(\C)$ to the category $\Vect$ of finite-dimensional complex vector spaces. The morphisms of $\threecob(\C)$ are cobordisms, i.e. three-manifolds with embedded ribbon graph. The TFT-construction provides the correlator as the invariant of such a cobordism.
The correlator of a topological world sheet $\WS$ is  an element in a finite dimensional vector space. This space can be identified with the space of conformal blocks on the world sheet $\WSg$, obtained by endowing $X$ with a complex structure. Thus the structure constants of the expansion of such a correlator are the same as the ones for the correlation function\footnote{The correlation function depends in general on the metric on $\WSg$. However, a certain quotient of correlators will only depend on the conformal equivalence class of the metric, see \cite[section 6.1.4]{TFTiv}. It is these quotients that can be obtained via the TFT-construction.}.

In section 2 we review some basic facts concerning modular tensor categories and the TFT-construction. Section 3 describes how the factorization constraints are implemented on a specific correlator. There are 2 types of factorization, bulk and boundary factorization. Boundary factorization is covered only briefly since we do not need it for the calculations in this paper. In section 4 we use bulk factorization to show how $\CA$ and its representation theory appear in the annulus partition functions.

\section{Modular tensor categories and the $\tft$-functor}
A modular tensor category $\C$ is in particular an abelian, semisimple, $\mathbb C$-linear, ribbon category. Thus any object is a finite direct sum of simple objects. Since the ground field of $\C$ is $\mathbb C$ the notion of a simple object is the same as a "scalar" object, meaning that $\End(U_i)=\mathbb C$.
We choose representatives of isomorphism classes of simple objects and label them by a finite index set $\I$, i.e.
\be
    \{U_i\,|\,i\in\I\},
\ee
where we take $U_0=\one$ and $\kb\in\I$ such that $U_\kb\cong U_k\v$ for all $k\in\I$. Since $\C$ is ribbon we make extensive use of graphical calculus, see e.g.\! \cite[section 2]{TFTi}. Due to strictness lines labeled by $\one$ are invisible. Among the structures defining a ribbon category is the twist. We denote the twist of the object $U$ by $\theta_U$. The twist of a simple object  $U_i$, which is proportional to the identity morphism, is written as
\be
    \theta_{U_i}=\theta_i\,\id_{U_i},\quad\theta_i\in\mathbb C.
\ee
In a modular tensor category there is also a non-degenerate matrix $S$, c.f. \cite[eqs. (2.21) \& (2.27)]{TFTi}, which is part of a representation of the modular group. We will also use the quantum dimension $\dim(U)$ of an object $U$, c.f. \cite[eq.\! (2.17)]{TFTi}, which for a simple object is related to the $S$-matrix:
\be
    \dim(U_i):=\frac{S_{i,0}}{S_{0,0}}.
\ee
\subsection{Algebras in tensor categories}

An algebra in a modular tensor category is an object $A$, equipped with a product $m\in\Hom(A\oti A,A)$ and a unit $\eta\in\Hom(\one,A)$ that satisfy associativity and unit constraints:
\be
    m\circ(\id_A\oti m)=m\circ(m\oti\id_A) \quad\text{ and }\quad m\circ(\eta\oti\id_A)=\id_A=m\circ(\id_A\oti\eta).
\ee
Similarly a coalgebra  $A$ in $\C$ is an object $A$, together with a coproduct $\Delta\in\Hom(A,A\oti A)$ and a counit $\eps\in\Hom(A,\one)$ satisfying coassociativity and counit constraints:
\be
    (\id_A\oti \Delta)\circ\Delta=(\Delta\oti\id_A)\circ\Delta \quad\text{ and }\quad (\eps\oti\id_A)\circ\Delta=\id_A=(\id_A\oti\eps)\circ\Delta.
\ee
A \emph{Frobenius} algebra in a tensor category is an object $A$ which is both an algebra and a coalgebra, such that the product and coproduct obey the following compatibility condition
\be
    (\id_A\oti m)\circ(\Delta\oti\id_A)=\Delta\circ m=(m\oti\id_A)\circ(\id_A\oti\Delta).
\ee

A left-module $M\equiv(M,\rho)$ over an algebra $A$ in $\C$ is an object $M$, equipped with a representation morphism $\rho\in\Hom(A\oti M,M)$ satisfying
\be
    \rho\circ(m\oti \id_M) =\rho\circ(\id_A\oti \rho)\quad\text{ and }\quad \rho\circ(\eta\oti \id_M)=\id_M.
\ee
Similarly a right-module over $A$ is an object $M$, together with a morphism\linebreak $\rho\in\Hom(M\oti A,M)$, satisfying analogous relations. For two algebras $A$ and $B$ in a tensor category, an $A$-$B$-bimodule $X\equiv(X,\rho_L,\rho_R)$ is an object $X$, such that $(X,\rho_L)$ is a left $A$-module and $(X,\rho_R)$ is a right $B$-module, such that the two actions commute. A simple module is a module
that does not have a non-trivial subobject which is a module itself.
For any two left $A$-modules $M$ and $N$, we define the subspace of left $A$-module morphisms
\be
    \Homa M N:=\{f\in\Hom(M,N)\,|\;\rho_N\circ (\id_A\oti f)=f\circ\rho_M\}.
\ee
Similarly, for any two $A$-$B$-bimodules $X$ and $Y$, the space
\be
    \Homab X Y
\ee
consists of all morphisms in $\Hom(X,Y)$ that commute with the left action of A and the right action of $B$. For any two objects $U$ and $V$ in $\C$ we define the $A$-$A$-bimodule $U\oti^+ A\oti^- V$ as
\be
\begin{split}
U\oti^+ A\oti^- V:=&\Big(U\oti A\oti V,\;\big[(\id_U\oti m\oti\id_V) \circ (c_{U,A}\inv\oti\id_A\oti\id_V)\big],\\
&\;\big[(\id_U\oti m\oti\id_V) \circ (\id_U\oti\id_A\oti c_{A,V}\inv)\big]\Big).
\end{split}
\ee
\subsection{The TFT-construction}
We review some aspects concerning the TFT-construction. A detailed description can be found in \cite[section 3-4]{TFTiv} or \cite{S}, see also  \cite[appendix A.1-A.5]{FSS} for a shorter description.

A modular tensor category $\C$ serves as a decoration of a geometric category $\threecob(\C)$. The objects of $\threecob(\C)$ are extended surfaces and the morphisms are cobordisms. An \emph{extended surface} $E$ is a  compact closed oriented two-manifold, with marked points and a choice of Lagrangian subspace $\lambda\subset H_1(E,\mathbb R)$. The data of a marked point contain in particular an object in $\C$. A \emph{cobordism} $\M\colon E\rightarrow E'$ is a compact oriented three-manifold, with boundary $\partial\M=(-E)\sqcup E'$ and an embedded ribbon graph with one ribbon ending at each marked point. The ribbon graph is colored by objects and morphisms in $\C$.

Given a modular tensor category $\C$ we can construct a three-dimensional topological field theory (3d TFT). A 3d TFT is a tensor functor from $\threecob(\C)$ to the category $\Vect$ of finite dimensional complex vector spaces.
Thus $\tft(E)\equiv\tfs(E)$ is a vector space and $\tft(\M)\equiv\tfm(\M)$ is a linear map
\be
\tfm(\M)\colon\tfs(E)\rightarrow\tfs(E').
\ee
By projecting a ribbon graph locally to $\mathbb R^2$ in a non-singular manner we can consider it as a morphism in $\C$ and manipulate the ribbon graph locally by the rules of graphical calculus. Transformations of this kind leave  the linear map $\tfm(\M)$ invariant.
Furthermore, the linear map $\tfm(\M)$ is a topological invariant and we will refer to it as the \emph{invariant} of $\M$.
A particular  extended surface is the double $\widehat\WS$ of a topological world sheet $\WS$. For the purposes of this paper we can identify the \tft-state space $\tfs(\widehat\WS)$ with the space of conformal blocks on $\widehat\WSg$.

The $\tft$-functor is central in the TFT-construction of rational CFT. The TFT-construction takes as input a modular tensor category $\C$ and (a Morita class\footnote{Morita equivalent algebras give rise to equivalent RCFT's.} of) a symmetric special Frobenius algebra $A$ in $\C$. These data define a unique RCFT. The TFT-construction provides the correlator of a world sheet $\WS$ by giving the construction of a cobordism $\M_\WS\colon\emptyset\rightarrow\widehat\WS$, the connecting manifold.
As a three-manifold, $\M_\WS$ is constructed by taking the interval bundle over $\WS$ and identifying points over the boundary:
\be\label{con_man}
    \M_\WS:=\WS\times[-1,1]\,\big/\sim\; ,\quad(x,t)\sim(x,-t)\;\forall \,x\in\partial \WSg \text{ and }\forall\, t\in[-1,1].
\ee
Thus $\partial \M_\WS\cong\widehat\WS$, c.f. \eqref{double}, and the world sheet is canonically embedded in $\M_\WS$ as all points in $(x,0)\in\M_\WS$.
Each field is indicated by a marked point on the world sheet $\WS$. A bulk fields gives rise to two marked points on $\widehat\WS$, c.f. \eqref{con_man}, whereas due to the identification of points over the boundary $\partial\WS$ in \eqref{con_man}, a boundary field gives rise to a single marked point on $\widehat\WS$.
The structure on the world sheet appears in $\M_\WS$ as parts of the ribbon graph. The boundary conditions are given by left $A$-modules and each boundary component appears as a ribbon, labeled by the corresponding $A$-module. We refer to a boundary condition labeled by a simple $A$-module as an elementary boundary condition. Field insertions appear as coupons, labeled by morphisms in $\Homaa{U\oti^+ A\oti^-V}A$ and $\Homa{M\oti U}M$, with appropriate objects $U$ and $V$, for bulk and boundary fields respectively.
The correlator $\Corr(\WS)$ is obtained from the invariant of the connecting manifold:
\be\label{corr_inv}
    \Corr(\WS)=\tfm(\M_\WS)\,1\in\tfs(\widehat\WS).
\ee
Since we identify $\tfs(\widehat\WS)$ with the space of conformal blocks on $\widehat\WSg$, \eqref{corr_inv} indeed defines a vector in the space of conformal blocks on $\widehat\WSg$.
For the rest of this paper we can and will make the identification
\be\label{corr_id}
    \Corr(\WS)\equiv\tfm(\M_\WS).
\ee

\section{The factorization constraints}

Factorization constraints  relate correlators of world sheets of (possibly) different topology. Starting from one world sheet, we can cut it along an embedded circle $S$, which results in two holes in the world sheet. A new world sheet $\WSf$ is obtained by gluing a half sphere, with one primary bulk field, to each hole. This describes \emph{bulk} factorization. \emph{Boundary} factorization amounts to cutting the world sheet along a line $\ell$ joining two boundary components,  closing the gaps in the boundary by gluing a half disc with a boundary field to each gap, and sum over all elementary boundary fields.

The correlators provided by the TFT-construction satisfy all factorization constraints \cite{TFTv}. We will restrict the discussion to orientable world sheets. The unorientable case works in a similar manner. 
Factorization is described in detail in  \cite[section 2]{TFTv}.

A factorization introduces extra field insertions on the world sheet $\WSf$, obtained after factorization. As a consequence, if the double $\widehat\WS$ is marked by $n$ points, the number of marked points on the double $\widehat\WSf$ of the new world sheet will be $n+2$ after boundary factorization and $n+4$ after bulk factorization. Thus $\tfs(\widehat\WSf)\ncong\tfs(\WS)$ and consequently, the correlator of the factorized world sheet $\WSf$ is not in the same space as the correlator of the original world sheet. The factorization constraints, satisfied by the correlators of the TFT-construction, are stated in \cite[theorem 2.9]{TFTv} (boundary factorization) and \cite[theorem 2.13]{TFTv} (bulk factorization). The theorems states first of all that there exists a \emph{gluing homomorphism}
\be
		G:\tfs(\widehat\WSf)\rightarrow\tfs(\widehat\WS).
\ee
The composition $G\circ\Corr(\WSf)$ is thus in the same space as $\Corr(\WS)$. Second, the two theorems show how these vectors are related. Schematically we can write this as
\be\label{fact_sch}
    \Corr(\WS)\sim\sum_{\text{fields}} G\circ\Corr(\WSf),
\ee
where the summation is over primary boundary fields or primary bulk fields depending on what kind of factorization we are considering.
For the purposes of this paper we do not need the gluing homomorphism explicitly. We rather need the action of the gluing homomorphism on some specific correlator. Remember \eqref{corr_inv} that the correlators are given by invariants of cobordisms. The gluing homomorphism $G$ is also given as an invariant of a cobordism
\be
\tilde G:\widehat\WSf\rightarrow\widehat\WS.
\ee
Let $\MXf\colon\emptyset\rightarrow\widehat\WSf$ be the connecting manifold of the factorized world sheet. The $\tft$-functor implies that there exists a cobordism $\Mfg=\tilde G \circ\MXf$ such that
\be\label{factMF}
		\tfm(\Mfg)=\tfm(\tilde G) \circ\tfm(\MXf)=G\circ\Corr(\WSf).
\ee

The proof of factorization is a local issue in the sense that it involves only the fibers over a small neighborhood of the circle $S$ or line $\ell$ along which the factorization is performed. Thus, for the proof, the explicit form of $\Mfg$ is not needed. This is also a strength of the proof: The factorization constraints should be satisfied for any number of factorizations. Since the proof of factorization is a local consideration it treats an infinite number of factorizations simultaneously. On the other hand, for actual calculations of the correlator of the factorized world sheet we need to know $\Mfg$ explicitly. Below we review how this manifold is constructed in the case of boundary and bulk factorization. We refer the reader to \cite{TFTv} for the proof.
\subsection{Boundary factorization}
Boundary factorization is a local issue also on the level of the connecting manifold. The cobordism $\Mfg$ is obtained by applying an equality of morphisms in $\C$ to $\M_\WS$. Consider a strip of the world sheet with boundary conditions labeled by the left $A$-modules $M$ and $N$. The ribbon graph in this neighborhood can be taken to be on a form that, when interpreted as a morphism in $\C$, is a certain projector $P_{M\v N}\in\End(M\v\oti N)$, c.f. \cite[eq.\! (4.7)]{S}. The manifold $\M_{q\gamma\delta}$, playing the role of $\Mfg$ in the case of boundary factorization, is then obtained by applying \linebreak\cite[eq.\! (4.22)]{TFTv} to $P_{M\v N}$, c.f.\! \cite[eq.\! (4.8)]{S}. The labels $\gamma$ and $\delta$ label the two boundary fields $\psi_\gamma\in\Homa {N\oti U_q} M$ and $\psi_\delta\in\Homa {M\oti U_{\qb}} N$ respectively. The invariant of $\M_{q\gamma\delta}$ is related to $\tfm(\MX)$ by
\be\label{Boundary_fact_inv}
		\tfm(\MX)=~\sum_{q\in\I}\sum_{\gamma,\delta}({c^{\text{bnd}}_{N,M,q}})
  \inv_{\;\delta\gamma}\,\tfm(\M_{q\gamma\delta}).
\ee
The elements of the matrix $({c^{\text{bnd}}_{N,M,q}})$ are the structure constants of the correlator of the disc with two boundary fields $\psi_\gamma$ and $\psi_\delta$, see \cite[eq. (2.27)]{TFTv}.
\subsection{Bulk factorization}
Bulk factorization is a more involved issue. The reason is that the construction of $\Mfg$ is a non-local problem.
Bulk factorization is performed along an embedding $\wsemb(S)$ of a circle $S$ in $\WS$. We will be interested in a millstone-shaped neighborhood of $\millstone_X\subset\M_X$ obtained as the fibers over a tubular neighborhood of $\wsemb(S)$. The preimage
\be
		Y_S:=\pi_X\inv(\wsemb(S))\in\MX,
\ee
of $\wsemb(S)$ under the canonical projection $\pi_\WS$ from $\MX$ to $\WS$ (c.f. \eqref{con_man}) separates $\millstone_X$ into two disjoint parts. $Y_S$ is an annulus whose two boundary components are contained in the boundary of $\M_\WS$.
Removing $Y_S$ from $\MX$ and taking the closure results in a manifold with corners, $\MXcut$. The boundary of $\MXcut$ contains two copies $Y_S^1$ and $Y_S^2$ of $Y_S$.

The manifold $\Mfg$ in \eqref{factMF} is constructed by composing $\MXcut$ with another manifold with corners. This manifold which we denote by  $\T_{q_1q_2\gamma\delta}$ is as a three-manifold $D\times S^1$:
\eqpic{stickyft}{270}{110}{
   \put(0,-5){
   \put(0,110)   {$ \T_{\!q_1q_2\gamma\delta} ~= $}
    \put(30,0){
    \scalebox{0.77}{
  \put(75,0)  {  \Includepic{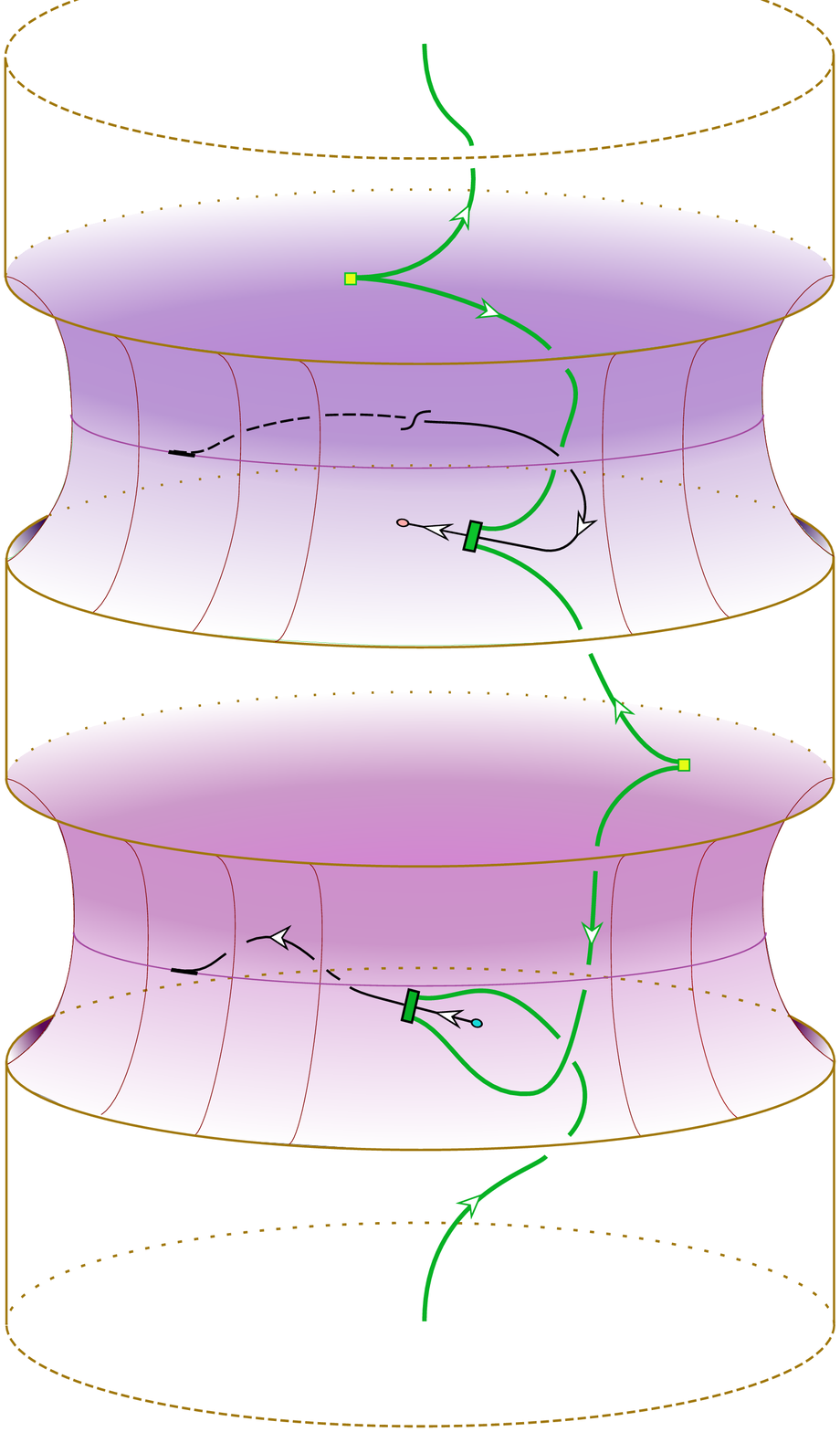} }
  \put(175,0) {
  \put(-5,26)     {\pg {q_2}}
  \put(14,215)    {\pg {\bar q_2}}
  \put(13,171)    {\pg {\bar q_1}}
  \put(36,129)    {\pg {q_1}}
  \put(-16,79)    {\pl {\phi_\gamma}}
  \put(-2,177.6)  {\pl {\phi_\delta}}
  \put(65,111)    {\lsqarrow}
  \put(65,219)    {\lsqarrow}
  \put(92,113)    {$\vio Y_\T^2 $}
  \put(92,221)    {$\vio Y_\T^1 $}
  } } } } }
Here $S^1$ is running vertically with top and bottom identified. We use black board framing for ribbon graphs, i.e. we depict ribbons as lines, see \cite[appendix. A.4]{FSS} for details.
The two spaces of bulk fields $\Homaa{U_{q_1}\oti^+A\oti^- U_{q_2}} A$ and $\Homaa{U_{\bar q_1}\oti^+A\oti^- U_{\bar q_2}}A$ are labeled by $\phi_\gamma$ and $\phi_\delta$ respectively. The boundary of $\T_{q_1q_2\gamma\delta}$ contains two copies of $Y_S$ as well. We denote them by $Y_\T^1$ and $Y_\T^2$. See \cite{FSS} for more details on $\T_{\!q_1q_2\gamma\delta}$.

The manifold $\M_{\WS;q_1q_2\gamma\delta}$, playing the role of $\Mfg$ in \eqref{factMF}, is obtained by identifying  $Y_S^{1}$ with $Y_\T^{1}$ and $Y_S^{2}$ with $Y_\T^{2}$.  There is a unique way to make this identification such that the orientations of the $A$-ribbons as well as the boundary components agree. The invariant of $\M_{\WS;q_1q_2\gamma\delta}$ is related to $\MX$ according to
\be    \label{bulkfact_Z}
\tfm(\MX)=\sum_{q_1,q_2\in\I}\sum_{\gamma,\delta}\dim(U_{q_1})\,\dim(U_{q_2})\,
  (c^{\text{bulk}}_{q_1,q_2})\inv_{\;\delta\gamma}\, \tfm(\M_{\WS;q_1q_2\gamma\delta}).
\ee
Here, $(c^{\text{bulk}}_{q_1,q_2})$ is a non-degenerate matrix whose elements are the structure constants of the two points function on the sphere, c.f. \cite[eq. (2.42)]{TFTv}
This is the precise form of \eqref{fact_sch} in the case of bulk factorization.
\section{The annulus partition functions}
Let the world sheet be an annulus with no field insertions, and with the boundary conditions at the two boundary components given by the simple $A$-modules $M$ and $N$ respectively.
The correlator of this world sheet is the annulus partition function  $\Aamp MN$, see \cite[section 5.8]{TFTi}. The double of the world sheet is a torus, and the connecting manifold $\M_{\Aamp MN}$ is a full torus with embedded ribbon graph, see \cite[eq.\! (5.117)]{TFTi}.
Consequently, the annulus partition function is an element in the space of conformal zero-point blocks on the torus. We choose a basis $\{\torchar k|k\in\I\}$ for this space with
\be\label{tor_bas}
    \torchar k=\tfm(\CMt k),\quad k\in\I.
\ee
The cobordism $\CMt k$ is a full torus with an annular ribbon labeled by $U_k$ inserted along the non-contractible cycle, c.f. \cite[eq.\! (5.15)]{TFTi}. The dual basis $\{\torchard k\;|k\in\I\}$ is given in \cite[eq.\! (5.18)]{TFTi}. The elements $\torchard k$ of the dual basis are obtained as
\be\label{tor_bas_dual}
    \torchard k=\tfm(\CMtd k),\quad k\in\I.
\ee
The manifold $\CMtd k$ differs from $\CMt k$ by reversion of the three-orientation and the orientation of the ribbon core, c.f. \cite[eq.\! (5.18)]{TFTi}. We wish to expand $\Aamp MN$ as
\be\label{exp_unfact}
    \Aamp MN=\sum_{k\in\I}\Aampk MNk\torchar k.
\ee
The duality of the bases implies that the annulus coefficients $\Aampk MNk$ are obtained by composing  $\CMtd k$ with $\M_{\Aamp MN}$.
This yields a ribbon graph in $S^2\times S^1$. $\Aampk MNk$ is obtained by applying the \tft-functor to this ribbon graph, c.f. \cite[section 5.8]{TFTi}.

\subsection{Factorization}
We investigate a bulk factorization along a circle $S$, embedded between and aligned with the two boundary components of the annulus. Using the prescription for bulk factorization, we first construct $\M_{\Aamp MN }^\circ$ by decomposing $\M_{\Aamp MN}$ into a disjoint sum of the following two components:
\eqpic{fact_ann1}{270}{27}{
   \put(20,25)   {$\M_{\Aamp MN }^{\circ,\;1}~=$}
   \put(80,-3){
   \scalebox{0.8}{
  \put(-7,-21)   {\Includepic{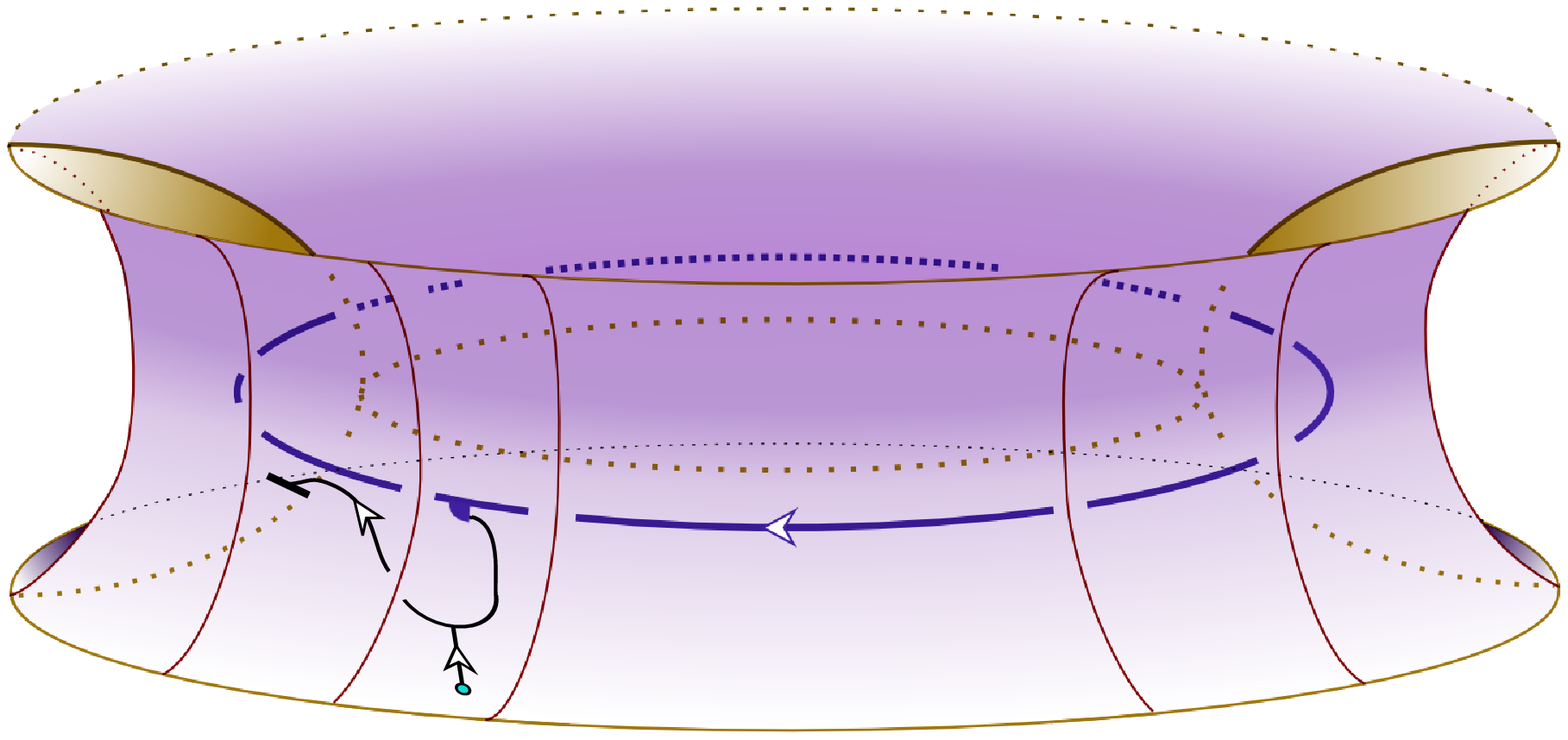}}
  \put(180,31)    {\begin{turn}{30}\lsqarrow\end{turn}}
  \put(207,47)    {$\vio Y_S^1 $}
  \put(110,11) {\pb M}
   } } }
   and
\eqpic{fact_ann2}{270}{25}{
    \put(20,25)   {$\M_{\Aamp MN }^{\circ,\;2}~=$}
    \put(80,-3){
   \scalebox{0.8}{
  \put(0,0)   {\Includepic{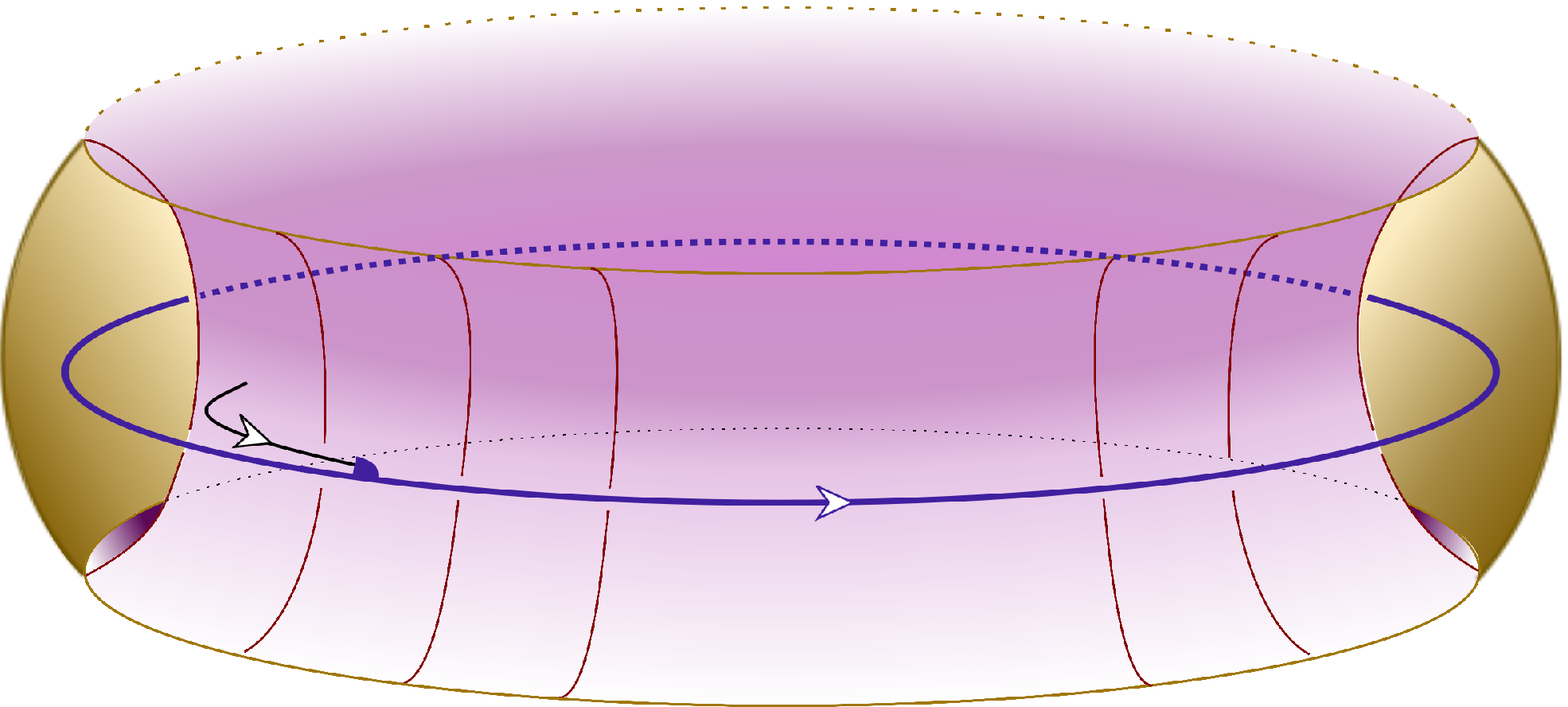}}
  \put(-12,99)    {\begin{turn}{-30}\longrsqarrow\end{turn}}
  \put(-25,100)    {$\vio Y_S^2 $}
  \put(95,20) {\pb N}
   } } }
Each component is a full torus with corners, with the boundary torus divided into two parts.
$Y_S^1$ and $Y_S^2$ constitutes the "outer" and "inner" part of the boundary of $\M_{\Aamp MN }^{\circ,\;1}$ and $\M_{\Aamp MN }^{\circ,\;2}$ respectively.  The remaining boundary parts constitute the boundary of $\M_{\Aamp MN}$.
The manifold $\M_{\WS;q_1q_2\gamma\delta}$ is obtained by composing $\M_{\Aamp MN }^\circ$ with $\T_{\!q_1q_2\gamma\delta}$. Following the prescription of the previous section we glue $\M_{\Aamp MN }^{\circ,\;1}$ and $\M_{\Aamp MN }^{\circ,\;2}$ to $\T_{\!q_1q_2\gamma\delta}$. The component $\M_{\Aamp MN }^{\circ,\;2}$
is readily composed with $\T_{\!q_1q_2\gamma\delta}$ by identifying $Y_S^2$ and $Y_\T^2$.

The composition of $\M_{\Aamp MN }^{\circ,\;1}$ with $\T_{\!q_1q_2\gamma\delta}$ is straightforward as well but needs a bit explanation.
First of all it has to be glued with the black side\footnote{ A ribbon with its preferred orientation is displayed as a solid line, whereas a dashed line, like the upper $A$-ribbon in \eqref{stickyft}, indicates that the ribbon is endowed with the opposite orientation. We refer to these to orientations as that the ribbon is showing its "white side" and its "black side" respectively.} of the ribbon graph facing upwards in order to match the $A$-ribbon in $\T_{\!q_1q_2\gamma\delta}$.
Second, think of $\M_{\Aamp MN }^{\circ,\;1}$ as a cylinder with the two opposite boundary discs identified, i.e. as $D\times[-1,1]$ with the discs $D\times\{-1\}$ and $D\times\{1\}$ identified. The composition is then performed by first identifying $Y_S^1$ with $Y_\T^1$ and afterwards identifying $D\times\{-1\}$ with $D\times\{1\}$.
The result is a cobordism $(\M_{\Aamp MN})_{\,q_1q_2,\gamma\delta}$, which is a ribbon graph in $D\times S^1$:
\eqpic{tstorus}{300}{100}{
\put(5,100)		{$(\M_{\Aamp MN})_{\,q_1q_2,\gamma\delta} ~= $}
\put(115,0){
\scalebox{0.8}{
\put(0,0){\Includepic{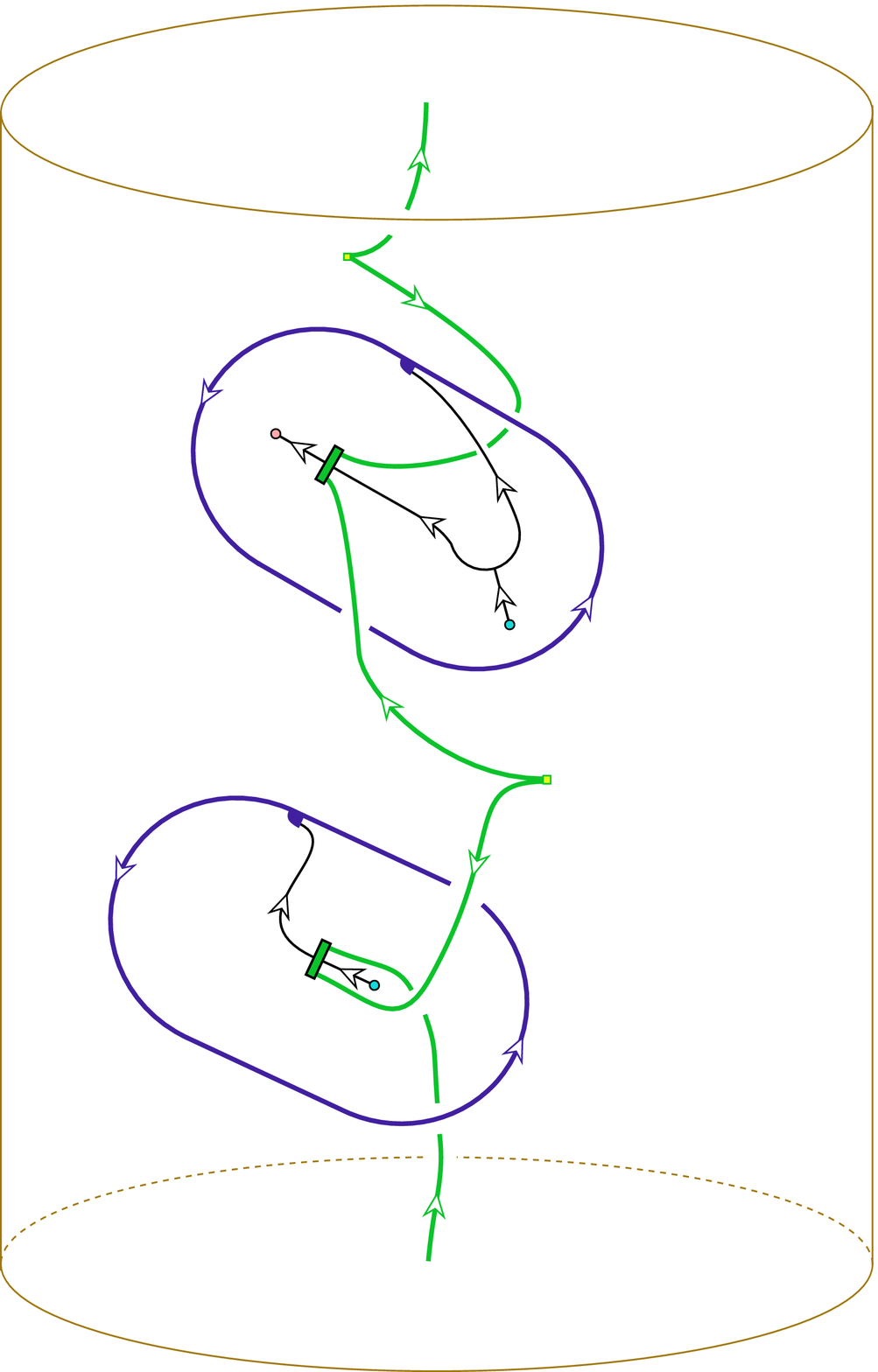}
\put(96,125)		{\pg {\bar q_1}}
\put(103,107)		{\pg {q_1}}
\put(92,32)		{\pg {q_2}}
\put(37,118)		{\pb{ N}}
\put(64,89)		{\pl{\phi_\gamma}}
\put(90,244)	{\pg {q_2}}
\put(91,214)	{\pg {\bar q_2}}
\put(66,188)	{\pl {\phi_\delta}}
\put(123,174)	{\pb{ M}}
}
}
}
}
Again $S^1$ is running vertically with top and bottom identified. Here we have also deformed the ribbon graph by a $\pi$ rotation of the part of the ribbon graph that shows its black side. The upper half of the ribbon graph can be interpreted as a morphism  in $\Hom(U_{\bar q_1},U_{q_2})$. This morphism can be non-zero only if $\bar q_1= q_2$. Consequently, the invariant is non-zero only if $q_1=\bar q_2$. Thus, applying \eqref{bulkfact_Z} the annulus partition function can be written as
\be\label{ann_amp_fact}
    \Aamp MN ~=~ \sum_{q\in\I}\sum_{\gamma,\delta=1}^{Z_{q\qb}}\dim(U_q)^2\,(c_{q,\qb}^{\text{bulk}^{-1}})_{\gamma\delta}\,\tfm((\M_{\Aamp MN})_{\,q\qb,\gamma\delta})\,.
\ee

\subsection{The annulus coefficients}
When an extended surface $E$ appears as the boundary of a three-manifold $M$, there is a canonical choice of Lagrangian subspace given by the kernel of the inclusion map $H(E,\mathbb R)\rightarrow H(M,\mathbb R)$. The purpose of the Lagrangian subspace is to define the surface unambiguously. Let $E$ be a torus and denote by the $A$-cycle the cycle that does not become contractible when $E$ appears as the boundary of a full torus, and let the $B$-cycle be the other one. The canonical choice of Lagrangian subspace of $H(\partial\M_{\Aamp MN},\mathbb R)$ and $H(\partial\CMtd k,\mathbb R)$ is spanned by the $B$-cycle. When we extract structure constants in \eqref{exp_unfact} by composing $\M_{\Aamp MN}$ and $\CMtd k$ we do this with the two B-cycles aligned.

The relation \eqref{bulkfact_Z} involves cutting out a full torus and gluing it back after an $S$-transformation. As a consequence, the factorization procedure exchanges the $A$- and $B$-cycles on  $\partial(\M_{\Aamp MN})_{\,q\qb,\gamma\delta}$ compared to $\partial \M_{\Aamp MN}$. Therefore, also the Lagrangian subspace is changed, such that in   \linebreak$H(\partial(\M_{\Aamp MN})_{\,q\qb,\gamma\delta},\mathbb R)$ it is spanned by the $A$-cycle. Thus, when extracting the annulus coefficient $\Aampk MNk$, after factorization, the manifold $\CMtd k$ has to be glued to $(\M_{\Aamp MN})_{\,q\qb,\gamma\delta}$ with the $B$-cycle on $\partial\CMtd k$ aligned with the $A$-cycle on $\partial (\M_{\Aamp MN})_{\,q\qb,\gamma\delta}$. The resulting cobordism $\MAnnk MNk q\gamma\delta$ is a ribbon graph in $S^3$:
\eqpic{tstorusS3}{300}{75}{
\put(40,75)		{$\MAnnk MNk q\gamma\delta~= $}
\put(110,-5){
\put(0,0){
\put(8,0) {\Includepic{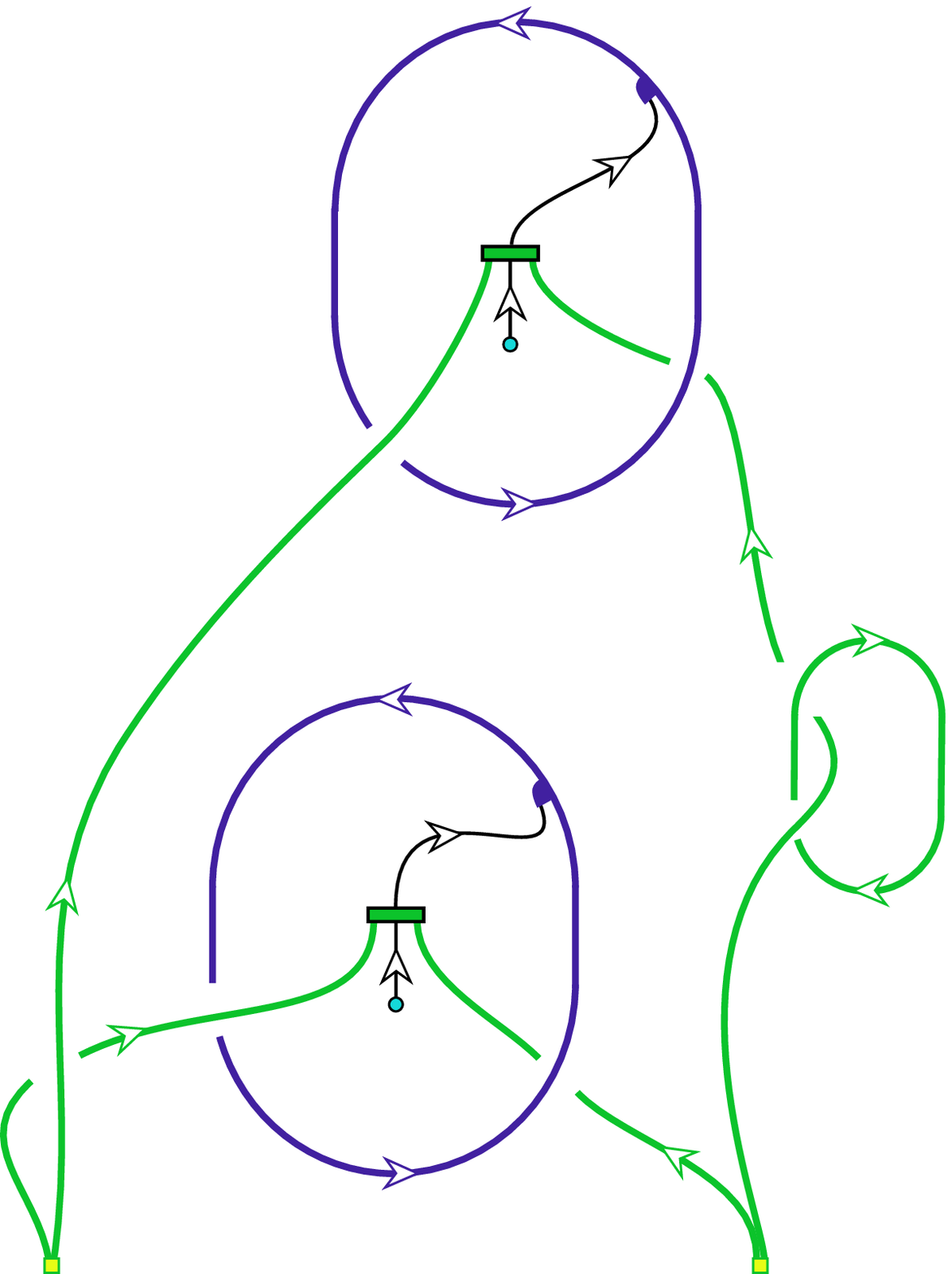}}
\put(24,15)		{\pg {\qb}}
\put(3,15)		{\pg {q}}
\put(92,12)		{\pg {\qb}}
\put(117,12)	{\pg {q}}
\put(37,73)		{\pb{ N}}
\put(68,47)		{\pl{\phi_\gamma}}
\put(110,70)	{\pg {k}}
\put(85,140)	{\pl {\phi_\delta}}
\put(98,169)	{\pb{ M}}
}
}
}
Combining with \eqref{ann_amp_fact}, the annulus coefficients can be written as
\be
    \Aampk MNk~=~\sum_{q\in\I}\sum_{\gamma,\delta=1}^{Z_{q\qb}}\dim(U_q)^2\,(c_{q,\qb}^{\text{bulk}^{-1}})_{\gamma\delta}\,\tfm(\MAnnk MNk q\gamma\delta).
\ee
In general, the choice of Lagrangian subspaces is related to an anomaly of the $\tft$-functor under gluing. However, in the case at hand the extended surfaces are doubles, which come with an orientation reversing involution. In this case this anomaly factor is unity, see \cite[Lemma 2.2]{FFFS}.

Next we evaluate $\tfm(\MAnnk MNk q\gamma\delta)$. The invariant of a ribbon graph in $S^3$ is calculated as follows: First we project the ribbon graph to $\mathbb R^2$ and interpret it as a morphism in $\C$. The result is an endomorphism of the tensor unit, i.e. a complex number. The invariant of the cobordism in $S^3$ is this number multiplied by $S_{0,0}$. Thus, we obtain after some manipulations
\be\label{ann_amp_RC}
    \Aampk MNk ~=~\dim(M)\dim(N)\sum_{q\in\I}S_{k,q}\theta_{q}\sum_{\gamma,\delta=1}^{Z_{q\qb}}
(c^{\text{bulk}}_{q\qb})^{-1}_{\delta\gamma}\,\refc q\gamma N\,\refc {\qb}\delta M\,.
\ee
The number $\refc q\gamma N$, with $q\in\I$ and $N$ a simple $A$-module, is a so-called reflection coefficients, c.f. \cite[eqs. (3.24) \& (3.26)]{S}. $\refc q\gamma N$ is related to the single structure constant, $\coronebd{\Phi_\gamma} N$, of the one-point correlator of the disc with boundary condition $N$ and a single bulk field, labeled by $\phi_\gamma\in\Homaa{U_q\oti^+A\oti^-U_{\qb}}A$, by $\coronebd{\Phi_\gamma} N=\dim (N)\,\refc q\gamma N$.
To arrive at the expression \eqref{ann_amp_RC} we first remove the annular $U_k$-ribbon, which yields a factor $\frac{S_{k,q}}{S_{0,0}}$. Second, we use dominance in $\End(U_{\qb}\oti U_q)$, which separates the morphism into two morphisms, each of them proportional to a reflection coefficient. Simplifying the results by braiding and fusion moves we arrive at \eqref{ann_amp_RC}.

The reflection coefficients can be calculated by evaluating the morphisms \linebreak\cite[eq.\! (3.24)]{S} in $\C$, corresponding to the one-point functions. However, they also appear as representation matrices of a semisimple associative complex algebra $\CA$\footnote{The complex algebra $\CA$ must not be confused with the Frobenius algebra $A$, which is an algebra in an abstract category $\C$.}, the classifying algebra \cite{FSS}. As a vector space, $\CA$ is given as the space of primary bulk fields with non-zero correlator on the disc:
\be
    \CA:=\bigoplus_{q\in\I}\Homaa{U_q\oti^+ A\oti^-U_{\qb}}A.
\ee
The irreducible representations of $\CA$ are all one-dimensional and are labeled by simple modules over $A$ in $\C$. Choosing a basis $\{\phi^{q,\alpha}|\alpha=1,...,Z_{q\qb}\}$ of $\Homaa{U_q\oti^+ A\oti^-U_{\qb}}A$, the representation matrices are $\rho_M(\phi^{q,\alpha})=\refc q\alpha M$.
Furthermore, $\CA$ is equipped with a non-degenerate bilinear form, $\nbfm$. In the basis $\{\phi^{q,\alpha}\}$, the bilinear form is given by $\nbfm(\phi^{p,\alpha},\phi^{q,\beta})=\nbf p\alpha q\beta$ where
\be\label{NBF}
    \nbf p\alpha q\beta=[\theta_p\dim(U_p)\,c_{00}^{\text{bulk}}]^{-1} \delta_{\bar q,p}\,
  c^{\text{bulk}}_{p\bar p,\alpha\beta}\,,
\ee
c.f. \cite[eq.\! (4.26)]{FSS}. $\nbfm$ is  a $\dim(\CA){\times}\dim(\CA)$ block matrix, where each block, labeled by $p\in\I$, is proportional to $c^{\text{bulk}}_{p\bar p}$.
Combining \eqref{ann_amp_RC} and \eqref{NBF}, we can rewrite the annulus coefficient $\Aampk MNk$ as
\be\label{ann_amp_BF}
    \Aampk MNk ~=~\frac{\dim(M)\dim(N)S_{0,0}^2}{\dim(A)}\sum_{q\in\I}\frac{S_{k,q}}{S_{0,q}}\sum_{\gamma,\delta=1}^{Z_{q\qb}}
\nbfinv {\qb}\delta q\gamma\,\refc q\gamma N\,\refc {\qb}\delta M\,.
\ee
Thus, much of the significant information concerning the annulus partition functions is contained in $\CA$ and its representation theory.

We conclude by comparing \eqref{ann_amp_RC} with some previous results. In the Cardy case, i.e. when $A$ is Morita equivalent to the tensor unit, the irreducible boundary conditions are labeled by simple objects $M{=}U_m$ and $N{=}U_n$ in $\C$. The matrix $(c^{\text{bulk}}_{q\qb})^{-1}_{\delta\gamma}$ is a scalar given by $\frac{S_{q,0}}{\theta_q}$, and the reflection coefficients are $\frac{S_{n,\qb}}{\dim(U_m)S_{q,0}}$ and $\frac{S_{m,q}}{\dim(U_m)S_{q,0}}$ respectively. Combining this with the Verlinde formula we obtain from \eqref{ann_amp_RC}
\be
    \Aampk mnk=\N kmn.
\ee
This  result was established already in \cite{Ca}, and it also follows directly from e.g.\linebreak \cite[eq.\! (2.16)]{BPPZ} or \cite[eq.\! (5.119)]{TFTi}.

In \cite[Theorem 5.20]{TFTi} some more results on the annulus coefficients are listed. The result \eqref{ann_amp_BF} corresponds to point (iv) in that list, with the difference that \eqref{ann_amp_BF} is written in a more symmetrical manner. Furthermore, using $S_{\qb,\kb}=S_{q,k}$ and $(c^{\text{bulk}}_{q\qb})^{-1}_{\delta\gamma}=(c^{\text{bulk}}_{\qb q})^{-1}_{\gamma\delta}$, the result \cite[Theorem 5.20 (iii)]{TFTi}
\be
\Aampk MNk=\Aampk NM\kb
\ee
is reproduced. Finally, combining \cite[Theorem 5.20 (ii)]{TFTi}, which states $\Aampk MN0=\delta_{M,N}$, with \eqref{ann_amp_BF}, we obtain an orthogonality relation for the representations of the classifying algebra:
\be
    S_{0,0}^2\sum_{q\in\I}\sum_{\gamma,\delta=1}^{Z_{q\qb}}
\mu\inv_{\qb\delta,q\gamma}\,\refc q\gamma N\,\refc {\qb}\delta M=\frac{\dim(A)}{\dim(M)\dim(N)}\,\delta_{M,N}\,.
\ee
\subsection*{Acknowledgements:} The author would like to thank J. Fuchs for helpful discussions.

\end{document}